\documentclass[12pt,epsfig,color]{article}
\usepackage{amsmath}
\usepackage{epsfig}
\usepackage{amssymb}
\usepackage{cite}
\input{epsf}
\def\beq{\begin{equation}}
\def\eeq{\end{equation}}
\def\beqa{\begin{eqnarray}}
\def\eeqa{\end{eqnarray}}

\def\beq{\begin{equation}}
\def\eeq{\end{equation}}
\def\beeq{\begin{eqnarray}}
\def\eeeq{\end{eqnarray}}
\def\to{\rightarrow}

\def\b0{b_0}

\begin{document}

   \DeclareGraphicsExtensions{.pdf,.gif,.jpg,.eps,.ps,.epsi}

\begin{titlepage}
\renewcommand{\thefootnote}{\fnsymbol{footnote}}
\begin{flushright}
%BNL-NT-04/41 \\
%RBRC-482 \\
% hep-ph/???
     \end{flushright}
\par \vspace{10mm}
\begin{center}
{\large \bf Helicity Parton Distributions from Spin Asymmetries \\[3mm]
in W-Boson Production at RHIC}
\end{center}
\par \vspace{2mm}
\begin{center}
{\bf Daniel de Florian${}^{\,1}$}
\hskip .2cm
and
\hskip .2cm
{\bf Werner Vogelsang${}^{\,2}$ }\\
\vspace{5mm}
${}^{1}\,$ Departamento de F\'\i sica, FCEyN, Universidad de Buenos Aires,\\
(1428) Pabell\'on 1 Ciudad Universitaria, Capital Federal, Argentina\\[2mm]
${}^{2}\,$ Institute for Theoretical Physics, Universit\"{a}t T\"{u}bingen, \\
Auf der Morgenstelle 14,  D-72076 T\"{u}bingen, Germany \\
%Physics Department and RIKEN-BNL Research Center, \\
%Brookhaven National Laboratory, Upton, NY 11973, U.S.A.\\

\end{center}

%%%%%%%%%%%%%%%%%%%%%%%%%%%%%%%%%%%%%%%%%%%%%%%%%%%%%%%%%%%%%%%%%%%%%%%%%%%%%%%%%%%%%%%%
%%%%%%%%%%%%%%%%%%%%%%%%%%%       ABSTRACT      %%%%%%%%%%%%%%%%%%%%%%%%%%%%%%%%%%%%%%%%
%%%%%%%%%%%%%%%%%%%%%%%%%%%%%%%%%%%%%%%%%%%%%%%%%%%%%%%%%%%%%%%%%%%%%%%%%%%%%%%%%%%%%%%%

\par \vspace{9mm}
\begin{center} {\large \bf Abstract} \end{center}
\begin{quote}
\pretolerance 10000
We present a next-to-leading order QCD calculation of the cross section
and longitudinal spin asymmetry in single-inclusive charged-lepton
production, $pp\to \ell^{\pm} X$, at RHIC, where the lepton is
produced in the decay of an electroweak gauge boson. Our calculation
is presented in terms of a multi-purpose Monte-Carlo integration 
program that may be readily used to include experimental
spin asymmetry data in a global analysis of helicity parton densities. 
We perform a toy global analysis, studying
the impact of anticipated RHIC data on our knowledge about the
polarized anti-quark distributions.

\end{quote}

\end{titlepage}

%%%%%%%%%%%%%%%%%%%%%%%%%%%%%%%%%%%%%%%%%%%%%%%%%%%%%%%%%%%%%%%%%%%%%%%%%%%%%%%%%%%%%%%%
%%%%%%%%%%%%%%%%%%%%%%%%%%%    INTRODUCTION     %%%%%%%%%%%%%%%%%%%%%%%%%%%%%%%%%%%%%%%%
%%%%%%%%%%%%%%%%%%%%%%%%%%%%%%%%%%%%%%%%%%%%%%%%%%%%%%%%%%%%%%%%%%%%%%%%%%%%%%%%%%%%%%%%

\section{Introduction}

Despite much progress over the past three decades, many open questions 
concerning the helicity structure of the nucleon remain. For 
example, we so far have only a rather sketchy picture of the individual 
polarizations of the light quarks and anti-quarks, $\Delta u/u, 
\Delta \bar{u}/\bar{u}, \Delta d/d, \Delta \bar{d}/\bar{d}$,
where the helicity parton distributions are as usual denoted by 
$\Delta q,\Delta \bar{q}$, and their spin-averaged counterparts by
$q,\bar{q}$. Since nucleons have up and down quarks as their valence 
quarks, the light-quark and anti-quark polarizations are of much interest in 
QCD and play key roles in many models of nucleon structure and, more
generally, for our fundamental understanding of the nucleon~\cite{dssvprd3C}.  

While lepton scattering has provided fairly precise and solid
information on the "total" up and down distributions, $\Delta u+\Delta\bar{u}$,
$\Delta d+\Delta\bar{d}$, through inclusive measurements, information 
on the individual $\Delta u, \Delta \bar{u}, \Delta d, \Delta \bar{d}$ is
much more sparse and still afflicted by large 
uncertainties~\cite{dssvprd,dssvprl}. The tool exploited here
so far has been semi-inclusive deep-inelastic scattering (SIDIS), 
in which one detects a specific hadron in the final state and
uses the fact that the flavor content of that hadron will typically be
correlated with the flavor of the quark or anti-quark hit by the virtual 
photon in the basic deep-inelastic reaction. Measurements of 
SIDIS spin asymmetries have vastly improved in recent years. By now,
quite precise data sets are available for various differently produced 
hadrons~\cite{ref:smc-sidis,ref:hermes-sidispd,ref:hermes-a1he3-sidisn,
ref:compass-sidisd,ref:compass-sidisnew}. On the other hand, extraction
of polarized parton distributions from SIDIS relies on the applicability of
a leading-twist factorized QCD description of the reaction, allowing among
other things the use of fragmentation functions \cite{dss} for the produced hadron determined from other processes. For the kinematics accessible in SIDIS 
so far one may worry if, for example, sub-leading twist effects can really
be ignored in the theoretical analysis of the data. This leads to an
uncertainty that is presently hard to quantify. It does have to be said,
however, that probably for the light quarks and anti-quarks, which are
primarily determined from pion SIDIS, this is less of a concern. Also, thanks
to the recent COMPASS measurements the kinematic reach of SIDIS data has 
become quite large now, extending into a regime where higher-twist
effects should be less relevant. In any case, as with any measurements
of nucleon structure, it is of great value to have a completely independent
probe that does not involve any hadronic fragmentation and is characterized
by momentum scales so large that perturbative calculations are expected to be
reliable and sub-leading twist effects unimportant.

It has long been recognized that $W^{\pm}$ boson production at the Relativistic
Heavy Ion Collider RHIC may provide unique and clean access to the individual
helicity polarizations of quarks and anti-quarks in the colliding protons~\cite{BouSof}. Thanks to maximal violation of parity in the elementary $Wq\bar{q}'$ 
vertex, $W$ bosons couple to left-handed quarks and right-handed
anti-quarks and hence offer direct probes of their respective helicity distributions
in the nucleon. Since spin asymmetries obtained from a {\it single} longitudinally
polarized proton beam colliding with an unpolarized beam are 
parity-violating for sufficiently inclusive processes, they have become 
the prime observables in the physics program with $W$ bosons at 
RHIC~\cite{BouSof,rhicrev,spinplan}:
\beq
A_{L}\equiv  \frac{d \sigma^{++}
+ d \sigma^{+-} - d \sigma^{-+} - d \sigma^{--}}{d \sigma^{++}
+ d \sigma^{+-} + d \sigma^{-+} + d \sigma^{--}}\equiv
\frac{d \Delta \sigma}{d \sigma} \; .
\label{eq:defl}
\eeq
Here the $\sigma^{++}$ etc. denote cross sections
for scattering of protons with definite helicities as indicated
by the superscripts. As one can see, the helicities of the second proton
are summed over, leading to the single-spin process $\vec{p}p\to W^{\pm}X$.
The basic idea behind measurements of the helicity distributions at RHIC is
then as follows: production of $W^-$, for example, selects 
primarily a $\bar{u}$ anti-quark from one proton in conjunction 
with a $d$-quark from the other. Thus, for the simple lowest-order (LO) 
parton-model process $d\bar{u}\to W^-$ the single-spin asymmetry becomes
\begin{equation}
A_{L}^{W^-}
\approx -\frac{\Delta d(x_1^0) \bar{u}(x_2^0) - \Delta \bar{u} (x_1^0) d (x_2^0) }
{d(x_1^0) \bar{u}(x_2^0) + \bar{u} (x_1^0) d (x_2^0) } ,
\label{loW}
\end{equation}
where $\Delta d,\Delta \bar{u}$ denote the usual helicity distributions, 
probed here at a scale of the order of the $W$ mass $M_W$, and where
\begin{equation}
x_{1,2}^0 = \frac{M_W}{\sqrt{S}}\, e^{\pm y_W}
\label{eq:x0def}
\end{equation}
with the rapdity $y_W$ of the $W$ bosons and the hadronic center-of-mass energy
$\sqrt{S}$. It follows that at large $y$, where
$x_1^0 \sim 1$ and $x_2^0\ll 1$, the asymmetry will
be dominated by the valence distribution probed at $x_1^0$,
and give direct access to $-\Delta d(x_1^0,M_W^2)/d(x_1^0,M_W^2)$. 
Likewise, for large negative $y$, $A_L^{W^-}$
is given by $\Delta \bar{u}(x_1^0)/\bar{u}(x_1^0)$. 
The situation for $W^+$ follows analogously. 

In practice, the above reasoning needs to be augmented in various 
ways. Foremost, there is an experimental issue: the detectors at RHIC
are not hermetic, which means that missing-momentum techniques
for the charged-lepton plus neutrino ($\ell\nu_\ell$) final states 
cannot be used to detect the $W$
and reconstruct its momentum. Instead, the strategy adopted by the 
RHIC experiments is to detect the charged decay lepton and
determine its transverse momentum $p_{T_l}$ and rapidity $\eta_l$. 
The relevant process therefore becomes the single-inclusive reaction
$pp\to \ell^{\pm} X$, similar in spirit to the processes $pp\to \pi X$, $pp\to
{\mathrm{jet}}X$~\cite{pionref,deFlorian:1998qp,jetref} 
used at RHIC to determine gluon polarization in the
nucleon. The ensuing expression for the single-spin asymmetry for
$pp\to \ell^{\pm} X$ differs from that in Eq.~(\ref{loW}) even at lowest 
order, since the lepton transverse momentum and rapidity do not 
completely determine the momentum fractions of the initial partons, so 
that an integration over momentum fractions appears in the expression.
Nevertheless, studies have shown~\cite{NadYuan1,NadYuan2,spinplan}
that despite this fact there should still be excellent sensitivity
to the distributions $\Delta u, \Delta \bar{u}, \Delta d, \Delta \bar{d}$ for
appropriately chosen lepton kinematics. 
Very recently, the RHIC collaborations
have presented the first preliminary data on the cross section and
single-spin asymmetry for $W^{\pm}$ boson production at 
RHIC~\cite{star,phenix}.

There are also theoretical issues that modify the simple picture given by 
Eq.~(\ref{loW}), regardless of whether one uses $W$ or lepton 
kinematics. Cabibbo-suppressed contributions, which involve the polarized 
and unpolarized strange quark distributions, and also contributions by
$Z$ bosons, are relatively straightforward to take into account.
A more important issue is the higher-order QCD corrections to the
leading-order (LO)
process $q\bar{q}'\to W^{\pm}$. At next-to-leading order (NLO), 
for example, one has the partonic reactions $q\bar{q}'\to W^{\pm}g$ and
$qg\to W^{\pm}q'$. Despite the fact that the $W$ mass sets a rather large
scale so that the strong coupling constant $\alpha_s(M_W)$ is not large,
the corrections can be significant and certainly need to be known for a 
reliable theoretical extraction of spin-dependent parton distributions from
data. Consequently, a lot of theoretical work has gone into the calculation 
of higher-order QCD corrections to the spin asymmetries in weak-boson 
production at RHIC. Early work in this area~\cite{ratcliffe,weber,weberqt,kamal,tgdy,gehrmann,grh} focused
on the case where the $W$ boson is detected directly, which is 
kinematically simpler and allows one to obtain analytical results for the 
NLO corrections. More recently, also an all-order resummation of 
terms in the partonic cross section that are logarithmically 
enhanced near partonic threshold was presented for this case~\cite{asmita}.
While, as we discussed above, the direct detection of the $W$ kinematics 
is not possible at RHIC, the relative size 
of the NLO corrections is expected to be rather independent of whether
one takes into account the $W\to \ell\nu_\ell$ decay or not, since this 
decay does not involve any strong interactions and all QCD corrections 
occur only in the initial partonic state. 

There have also been extensive studies of higher-order QCD corrections 
for the experimentally more relevant case of single-inclusive lepton production,
$pp\to \ell^{\pm} X$. In Refs.~\cite{NadYuan1,NadYuan2,rhicbos} the program
RHICBOS was introduced. RHICBOS is a Monte-Carlo integration program for
lepton distributions, specifically adapted to the polarized $pp$ collisions at RHIC. 
It puts particular emphasis on the effects of soft-gluon emission and their impact
on the region when the produced intermediate vector boson has small transverse
momentum, $q_T$. In the lowest-order diagrams $q\bar{q}'\to W\to \ell\nu_\ell$,
one has $q_T=0$. Gluon radiation generates a recoil transverse momentum. 
When $q_T$ tends to zero, large logarithmic corrections develop in the $q_T$ 
spectrum of the $W$'s. These can be resummed to all orders in perturbation
theory, following the Collins-Soper-Sterman (CSS) formalism~\cite{CSS}, 
which is done in RHICBOS at next-to-leading logarithmic level. 
%As such, 
%the RHICBOS program provides the NLO result, augmented by the all-order sum %of leading logarithmic terms. 
RHICBOS is widely used for phenomenological studies
related to the $W$ program at RHIC (see, for example, Ref.~\cite{spinplan}). 

Despite this earlier work,  
we will present in the present paper a new NLO calculation
of the cross sections and spin asymmetries for $pp\to \ell^{\pm} X$ at RHIC.
There are several reasons why this is in our view a necessary 
addition. First of all, it is of value to have an independent calculation 
of the relevant observables at RHIC. Second, the RHIC data
for the spin asymmetry in $W$ production will ultimately 
need to be included in a global NLO analysis of parton distributions
that includes all available information from lepton scattering and
$pp$ collisions at RHIC. Only then can the best possible 
information on the $\Delta q$ and $\Delta \bar{q}$ be extracted.
Inclusion of $pp$ scattering data in a global analysis is a relatively
complex task~\cite{dssvprd} since the computation of the parton 
subprocess cross sections is typically numerically quite involved. 
Recent papers~\cite{dssvprl,dssvprd} used a technique based on 
Mellin moments~\cite{sv} to achieve the first global analysis 
of polarized lepton scattering data and data for $pp\to \pi X$ and 
$pp\to{\mathrm{jet}}X$ from RHIC. Our calculation presented 
in this paper is set up in such a way that inclusion of RHIC data 
for $pp\to \ell^{\pm} X$ will be straightforward. This is an 
advantage over RHICBOS, which requires prior computations 
of certain "grid" files for a given set of parton distribution functions, 
and hence is to our knowledge not readily suited for use in a global  
analysis code. Using our new code, 
we will present in this paper a first "toy" study of 
a global analysis that includes projected or estimated data for $W$ 
observables at RHIC.

Finally, we also have a more theory-related reason for performing a 
new NLO computation of the $W$ observables at RHIC. As described
above, RHICBOS includes the all-order resummation of large 
logarithmic corrections arising at small $W$ transverse momentum $q_T$. 
As is well-known~\cite{tev}, these logarithms are very relevant if one is interested, 
for example, in the low-transverse momentum distribution of the $W$ boson itself. 
However, for the single-inclusive lepton cross section, the situation 
is somewhat different. $q_T$-resummation is really only useful when the 
observable is {\it directly} sensitive to (small) $q_T$. For
$pp\to \ell^{\pm} X$ this is the case when the measured lepton transverse 
momentum, $p_{T_l}$, is in the vicinity of $M_W/2$. This may be understood
as follows: for the LO reaction $q\bar{q}'\to W\to \ell\nu_\ell$, the lepton 
transverse momenta are basically limited to $p_{T_l}\leq M_W/2$, except for 
effects related to the finite decay width of the $W$. This means that lepton
transverse momenta $p_{T_l}> M_W/2$ primarily arise from higher-order 
gluon emission. Just above $M_W/2$, the lepton transverse momentum 
spectrum is then dominated by the same logarithms that are
present in the $W$ transverse
momentum distribution, which are resummed by the CSS formalism. This is
the motivation behind the resummation implemented in RHICBOS.
In practice, however, the RHIC experiments sample over a fairly 
broad range of $p_{T_l}$. For the theoretical
calculation this implies integration of the observables over $p_{T_l}$ over this 
range. The broader this range, the less dominant are the soft-gluon effects 
addressed by resummation, and the less useful is resummation. This becomes
particularly evident for the rapidity-distribution of the lepton, integrated over
$p_{T_l}$, which is the most relevant observable in $W$-physics at RHIC 
and, in the spirit of Eq.~(\ref{loW}), the best tool to separate the various
polarized parton distributions. To state our point more succinctly: for
practical purposes at RHIC, the region $p_{T_l}\approx M_W/2$ is only
a relatively small part of the sampled kinematics, so that the $q_T$ logarithms 
are not expected to dominate the observables. Their resummation is, then, 
not really appropriate and does not necessarily lead to an improvement of the 
theoretical calculation, since at the level of $p_{T_l}$-integrated observables 
there will be other higher-order effects that are of the same size as those provided 
by the terms logarithmic in $q_T$~\cite{joint,CdFG,Bozzi:2008bb}. In any case, from
the point of view of extracting polarized parton distributions, it is advisable 
in our view to select observables at RHIC that are insensitive to the 
complications associated with multiple soft-gluon emission.
It therefore seems preferable to us to use a plain 
NLO calculation for studies of $W$-production at RHIC, which we aim to do 
in this paper. 

The remainder of this paper is organized as follows:  in the next section
we very briefly discuss our NLO calculation, which is 
overall quite standard. The main part of the paper is then phenomenological 
and presented in the following two sections. Apart from providing NLO predictions 
for RHIC in Sec.~\ref{secIII}, we also present in Sec.~\ref{secIV} a 
"proof-of-principle" study of a global analysis involving $W$ spin asymmetries. 
We finally conclude in Sec.~\ref{secV}.

%%%%%%%%%%%%%%%%%%%%%%%%%%%%%%%%%%%%%%%%%%%%%%%%%%%%%%%%%%%%%%%%%%%%%%%%%%%%%%%%%%%%%%%%
%%%%%%%%%%%%%%%%%%%%%%%%%%%    Section  2       %%%%%%%%%%%%%%%%%%%%%%%%%%%%%%%%%%%%%%%%
%%%%%%%%%%%%%%%%%%%%%%%%%%%%%%%%%%%%%%%%%%%%%%%%%%%%%%%%%%%%%%%%%%%%%%%%%%%%%%%%%%%%%%%%

\section{Next-to-leading Order Calculation \label{secII}}

\begin {figure}[!ht]
\begin{center}
\includegraphics[width = 6.0in]{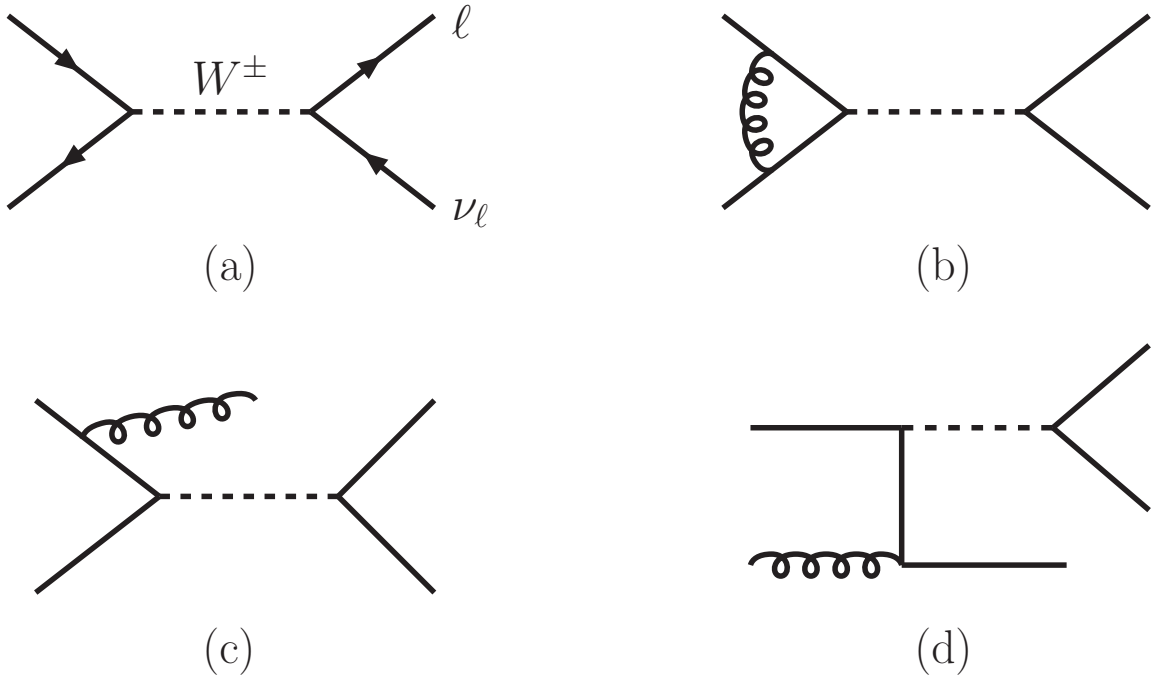} 
\end{center}%}
\caption{{\it Feynman diagrams for $W$ production with leptonic decay:
(a) leading-order, (b) NLO virtual correction, (c) NLO real emission,
(d) NLO quark-gluon scattering. Crossed diagrams are not shown.}
\label{figWdiag} }
\end{figure}

In order to evaluate the NLO QCD corrections to the process we rely on the 
version of the subtraction method introduced and extensively discussed in 
Refs. \cite{Frixione:1995ms,Frixione:1997np}, and later extended to the polarized case 
in Ref. \cite{deFlorian:1998qp}. We refer the reader to those references for the details. 
Figure~\ref{figWdiag} shows some of the Feynman diagrams contributing
at LO and NLO in the case of W exchange.
%%%%%%%%%%%%%%%%%%%%%%%%%%%%%%%%%%%%%%%%%%%%%%%%%%%%%%%%%%%%%%%%%%%%%%%%%%
%%%%%%%%%%%%%%%%%%%%%%%%%%%%%%%%%%%%%%%%%%%%%%%%%%%%%%%%%%%%%%%%%%%%%%%%%%%%%%%%
The calculation is implemented in the Monte-Carlo like code 'CHE' 
(standing for 'Collisions at High Energies')
\footnote{The code is available upon request from deflo@df.uba.ar} which 
provides access to the full kinematics of the final-state particles,  allowing
the computation of any infrared-safe observable in hadronic collisions and the 
implementation of realistic experimental cuts.
It is worth noticing that the same
code can compute the unpolarized, the single polarized and the double polarized
cross sections.

Besides the contribution driven by $W$ exchange, the code also allows the computation
of the background arising from $Z$-boson and/or photon exchange at the
same accuracy in perturbative QCD.
We point out that at NLO the contribution from photon exchange,
$q\bar{q}\to \gamma^* g$ followed by $\gamma^*\to \ell^+\ell^-$,
may generate large contributions when the high-transverse momentum
photon splits almost collinearly into the lepton pair, producing 
high-$p_{T_l}$ leptons with a very low invariant mass. A proper
treatment of this configuration would require the addition of a 
fragmentation contribution based on parton-to-dilepton fragmentation 
functions~\cite{Kang:2008wv}. On the other hand, it is likely that
configurations with two nearly collinear leptons can be distinguished
experimentally from true single high-$p_{T_l}$ leptons. 
In our calculation we avoid such configurations by requiring 
the lepton pair to have an invariant mass $M_{{l_1}{l_2}}>10$ GeV.

We note that we have checked the results for the spin-averaged
cross section in our code against the MCFM ~\cite{mcfm} and DYNNLO ~\cite{Catani:2009sm} codes. 
We have also computed the fully inclusive spin-averaged and 
polarized $W$ cross sections, integrated over all lepton angles. 
For these cross section, analytical results are 
available~\cite{kamal,tgdy,gehrmann}, with which we agree.

A virtue of our code is that it lends itself to inclusion in a global 
analysis of polarized parton distributions, along the lines presented
in Refs.~\cite{dssvprd,dssvprl}. In these papers, a method based
on Mellin moments was used~\cite{sv}, for which the theoretical 
expression for any cross section is split up into parts that are 
independent of the parton distributions, coupled to 
the Mellin moments of the parton distributions.
This procedure was shown to tremendously speed up the NLO fit, 
since the pieces that do not depend on the parton 
distributions, which contain the most time-consuming computations,
can be calculated "once and for all" prior to the fit and stored as large 
arrays. In the actual fit one then only needs to perform numerical
inverse Mellin transforms, which is straightforward. As was shown in 
Ref.~\cite{dssvprd}, the computation of the pre-calculated factors is
possible in a timely manner even for a code based on Monte-Carlo 
integration, provided a proper importance-sampling is used. We
have implemented the corresponding strategies described in~\cite{dssvprd} 
in our code.

%%%%%%%%%%%%%%%%%%%%%%%%%%%%%%%%%%%%%%%%%%%%%%%%%%%%%%%%%%%%%%%%%%%%%%%%%%%%%%%%%%%%%%%%
%%%%%%%%%%%%%%%%%%%%%%%%%%%    Section  3       %%%%%%%%%%%%%%%%%%%%%%%%%%%%%%%%%%%%%%%%
%%%%%%%%%%%%%%%%%%%%%%%%%%%%%%%%%%%%%%%%%%%%%%%%%%%%%%%%%%%%%%%%%%%%%%%%%%%%%%%%%%%%%%%%

\section{Phenomenological Results for RHIC \label{secIII}}

\begin {figure}[!ht]
\begin{center}
\includegraphics[width = 5.0in]{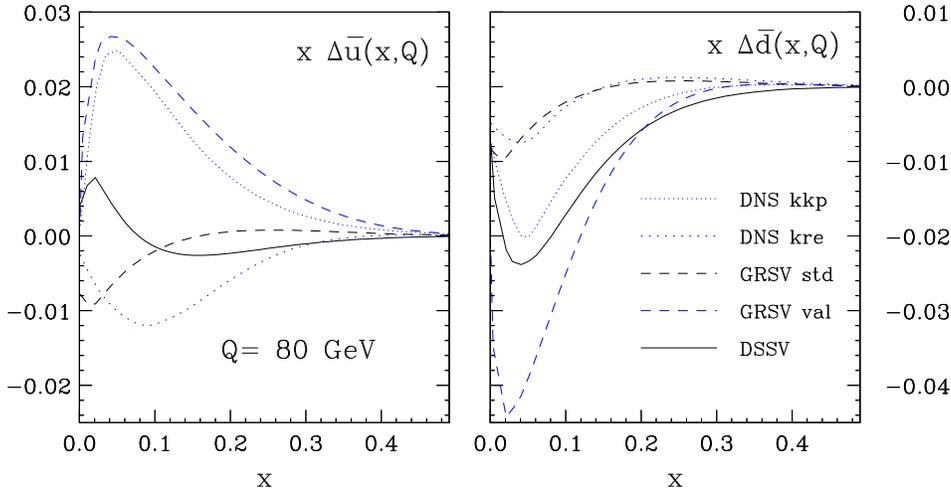} 
\end{center}%}
\caption{{\it Left: Next-to-leading order 
$x\Delta \bar{u}(x,Q)$ evaluated at the scale $Q=80$~GeV for the DSSV~\cite{dssvprd,dssvprl} (solid), GRSV~\cite{GRSV} (dashed), 
DNS-Kretzer~\cite{DNS} (dotted) and DNS-KKP (short-dashed) sets 
of polarized pdfs. Right: Same for $x\Delta \bar{d}(x,Q)$ (right-hand side)  } 
\label{figpdf} }
\end{figure}

We now use our NLO code to present some numerical results for polarized
$pp$ collisions at RHIC at center-of-mass energy $\sqrt{S}=500$ GeV. 
Our default choice for the spin-dependent parton distribution functions 
is the DSSV set~\cite{dssvprd,dssvprl}. 
Since we want to study the sensitivity of different 
observables to the polarized parton distributions, we will also consider a few other 
(and less recent) sets of polarized densities that primarily differ in the anti-quark polarizations: the "standard" and "valence" sets from GRSV \cite{GRSV}, 
which have SU(2) symmetric and broken sea distributions, respectively, and the 
"DNS-Kretzer" and "DNS-KKP" sets~\cite{DNS}. The last two sets correspond 
to fits to the same data for inclusive and semi-inclusive lepton scattering, but
obtained using different sets of fragmentation functions~\cite{Kretzer,KKP} 
to analyze the semi-inclusive asymmetries. We note that not all of these
additional sets of polarized parton distributions are completely compatible
with all information now available from SIDIS. However, given the 
potential uncertainties in SIDIS mentioned in the Introduction, they are useful
in order to gauge the sensitivity of future RHIC measurements. 
Figure~\ref{figpdf} shows a comparison of the light anti-quark distributions
$x\Delta \bar{u}(x,Q)$ and $x\Delta \bar{d}(x,Q)$ for the various sets,
evaluated at the scale $Q=80$ GeV relevant for $W$ production. 
As one can see, there are large differences among the distributions,
both in qualitative behavior regarding breaking of flavor-SU(2) symmetry,
and in magnitude. Since all sets provide very similar results for the sum 
of quark and anti-quark polarized distributions, $\Delta q_i+\Delta \bar{q_i} $, 
and since the observables are rather insensitive to the polarized gluon density, 
differences in the $W$ spin asymmetries computed with the different sets 
can be mostly attributed to the differences in the sea distributions. 

%%%%%%%%%%%%%%%%%%%%%%%%%%%%%%%%%%%%%%%%%%%%%%%%%%%%%%%%%%%%%%%%%%%%%%%%%%
%%%%%%%%%%%%%%%%%%%%%%%%%%%%%%%%%%%%%%%%%%%%%%%%%%%%%%%%%%%%%%%%%%%%%%%%%%%%%%%%

In order to compute the unpolarized cross section in the denominator 
of the spin asymmetries, we use the MRST2002 \cite{mrst2002} NLO set. 
This choice is motivated by the fact that the MRST2002 set was also   
used as the "baseline" unpolarized set for the DSSV~\cite{dssvprd,dssvprl} 
spin-dependent 
parton distributions. We have verified that the use of more recent sets of 
distribution functions, like CTEQ6~\cite{cteq6}, results in very similar cross 
sections. 

We set the masses of the vector bosons to $M_Z=91.1876$~GeV and
$M_W=80.398$~GeV, and the corresponding decay widths to 
$\Gamma_Z=2.4952$~GeV and $\Gamma_W=2.141$~GeV~\cite{PDG}. 
We neglect contributions from $b$ and $t$ quark initial states to 
$W$ production and, consistent with that, use the following values for the 
Cabibbo-Kobayashi-Maskawa (CKM) matrix elements: $|V_{ud}|=|V_{cs}|=
0.975$ and $|V_{us}|=|V_{cd}|=0.222$. We do not include any QED or 
electroweak corrections, but choose the coupling constants $\alpha$ and 
$\sin^2\theta_W$ in the spirit of the "improved Born 
approximation"~\cite{IBA1,IBA2}, in order to effectively take into account 
the electroweak corrections. This approach results in $\sin^2\theta_W=0.23119$
and $\alpha=\alpha(M_Z)=1/128$. We also require the lepton pair to have an invariant mass $M_{{l_1}{l_2}}>10$ GeV, in order to avoid potentially 
large NLO contributions from production of a high-$p_T$ nearly real photon 
that subsequently decays into a pair of almost collinear leptons, as discussed
in Sec.~\ref{secII}.

We will study two different observables for lepton production in
$pp\to \ell^{\pm}X$: the 
transverse momentum ($p_{T_l}$) distribution of the charged lepton 
with a rapidity cut of $|\eta_l|<1$, and the rapidity distribution with 
$p_{T_l}>20$ GeV. We count rapidity as positive in the
{\it forward} direction of the polarized proton. 
There are two hard scales in the process, which are of the same order:
the mass of the gauge boson and the transverse momentum of the observed 
lepton. We choose $\mu_F^2=\mu_R^2=(M_W^2+p_{T_l}^2)/4$ as the 
default factorization and renormalization scales for both the 
$W$ and the $Z/\gamma$ contributions. We note that the scale
dependence of the cross sections and, in particular, the spin asymmetries
is extremely mild in case of vector boson production, so that other choices like
$\mu_F=\mu_R=M_W$ give very similar results.

%%%%%%%%%%%%%%%%%%%%%%%%%%%%%%%%%%%%%%%%%%%%%%%%%%%%%%%%%%%%%%%%%%%%%%%%%%
\begin {figure}[!ht]
\begin{center}
\includegraphics[width = 5.0in]{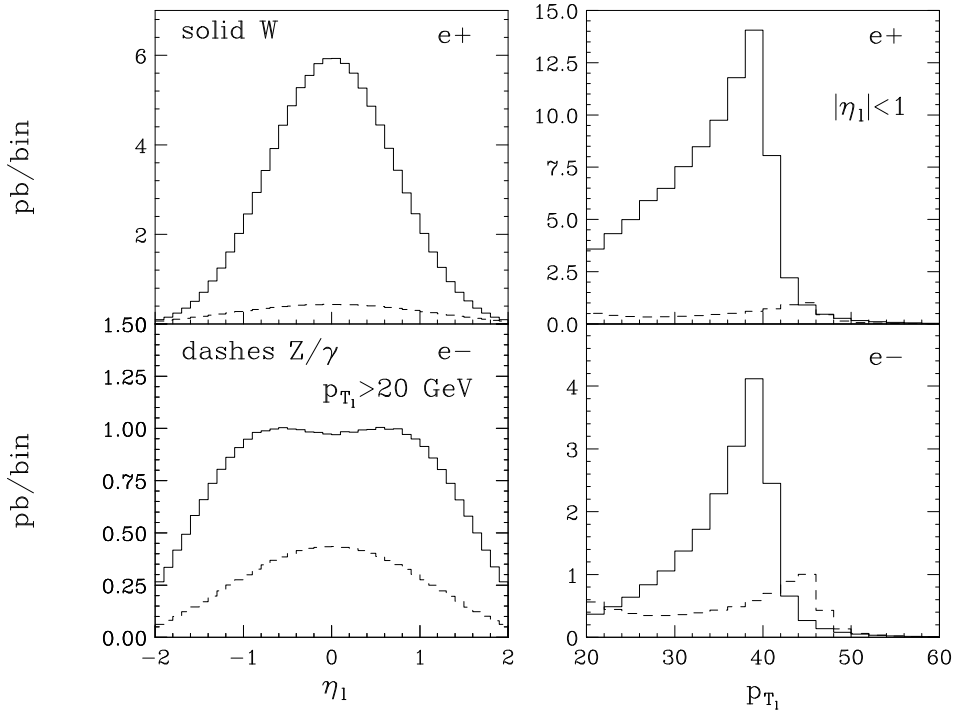} 
\end{center}%}
\caption{{\it Contribution from $W$ (solid lines) and $|Z|^2+\gamma Z+|\gamma|^2$ (dashed lines) production to the rapidity (left-hand side) and transverse momentum 
(right-hand side) distributions of the leptons in spin-averaged collisions
at RHIC. The upper plot corresponds to positron and the lower plot to electron production in unpolarized collisions.}
\label{figWZunpol} }
\end{figure}
%%%%%%%%%%%%%%%%%%%%%%%%%%%%%%%%%%%%%%%%%%%%%%%%%%%%%%%%%%%%%%%%%%%%%%%%%%%%%%%%

We start by investigating the contribution by $Z$ and $\gamma$ exchange to the cross-sections. Even though the $q\bar{q}Z$ coupling is also parity-violating
and hence may contribute to the single-longitudinal spin asymmetry, 
the $Z$ and $\gamma$ contributions are to be regarded in a sense as 
"dilutions" of the $W$ signal. Being almost symmetric in $\ell^+$ and $\ell^-$, 
they will somewhat decrease the clear-cut sensitivity of $A_L$ to the polarized 
sea-quark distributions, and they also contribute to the spin-averaged cross
section. Figure~\ref{figWZunpol} compares the $Z/\gamma$ contributions (dashes) to the ones arising from $W$ (solid), for the spin-averaged cross section as functions of rapidity (left-hand side) and transverse momentum 
(right-hand side). As can be observed, for positrons (upper row) the contribution from $Z$ turns out to be rather small in the central rapidity range (about 
$\sim 7\%$) while it does significantly add for electrons (lower row), reaching 
more than $40\%$ of the $W^-$ contribution. The effect turns out to be more noticeable at larger rapidities. We note that the $|Z|^2$ 
contribution dominates strongly over the $\gamma Z$ interference and
$|\gamma|^2$ ones. 
In the case of the transverse momentum distribution, the relative contribution
by $Z$'s strongly depends on $p_{T_l}$. Since the peak of the distribution occurs around $p_{T_l} \sim M_V/2$, with $M_V$ the vector boson mass, 
the difference between the $W$ and $Z$ masses induces dominance 
of the $Z$ contribution for $p_{T_l}\gtrsim 45$ GeV. 

%%%%%%%%%%%%%%%%%%%%%%%%%%%%%%%%%%%%%%%%%%%%%%%%%%%%%%%%%%%%%%%%%%%%%%%%%%
\begin {figure}[!ht]
\begin{center}
\includegraphics[width = 5.0in]{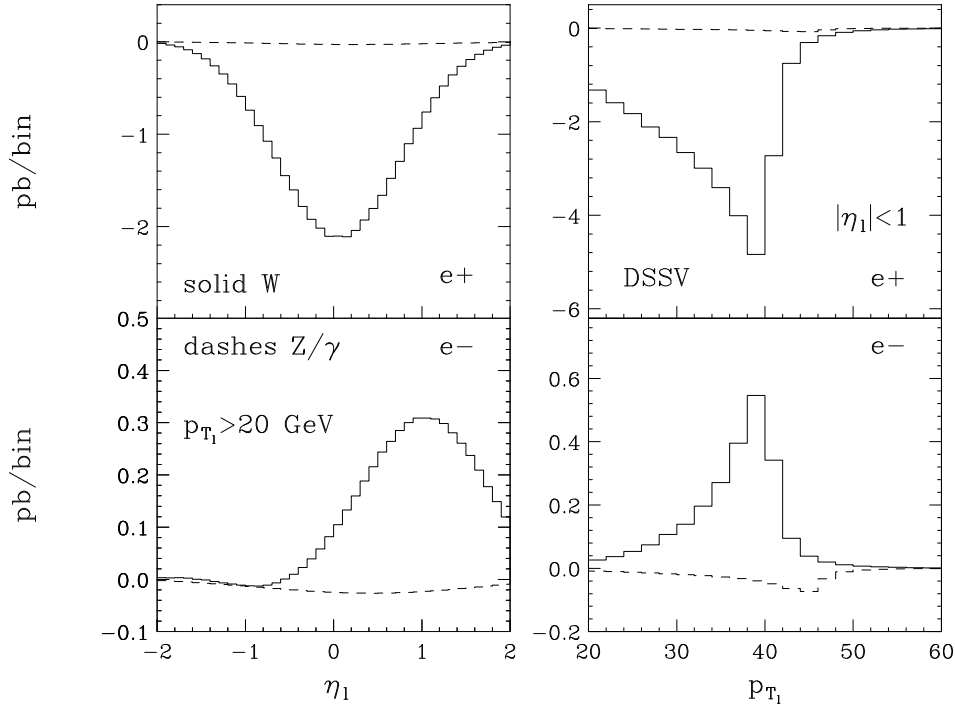} 
\end{center}%}
\caption{{\it Same as Fig.\ref{figWZunpol}, but for single-polarized collisions 
using the DSSV parton distributions.}
\label{figWZpol} }
\end{figure}
%%%%%%%%%%%%%%%%%%%%%%%%%%%%%%%%%%%%%%%%%%%%%%%%%%%%%%%%%%%%%%%%%%%%%%%%%%%%%%%%

The same comparison is shown in Fig.~\ref{figWZpol} for the single-polarized case, where we rely on the DSSV set of parton distributions for the polarized beam. While $Z$ exchange can produce a single-spin cross-section, its parity violation component is rather small and results in a contribution that does not exceed a few percent 
of the dominant $W$ one. Of course, this does not mean that 
the $Z$ contribution can be neglected in the analysis to be carried out to extract the polarized parton distributions at RHIC. The main observable is the spin asymmetry which, at least in the case of electron production, can be considerably reduced by the $Z/\gamma$ contribution in the unpolarized cross section. Vetoing events with a lepton pair could be valuable in this case in order to increase the sensitivity to the polarized parton densities. Conversely, $Z$ bosons by themselves may offer
an interesting advantage if both charged leptons from 
the decay $Z\to \ell^+\ell^-$ can be detected, because in that case one
can in principle reconstruct the kinematics of the $Z$ boson, which is
not possible for the $W$s because of the neutrino in its decay. One would then be
able to directly access the momentum fractions of the parton distributions,
in the spirit of Eq.~(\ref{loW}). Unfortunately, statistics for reconstructed 
$Z$ decays with both decay leptons will likely remain rather low at RHIC. 

\begin {figure}[!ht]
\begin{center}
\includegraphics[width = 5.0in]{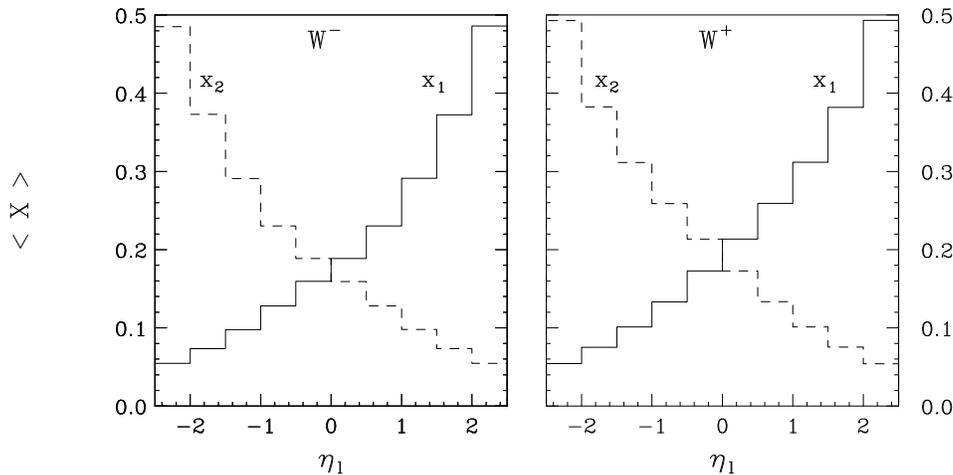} 
\end{center}%}
\caption{{\it Averages of the momentum fractions $x_{1,2}$ 
as functions of the charged lepton's rapidity $\eta_l$ 
for $W^-$ (left) and $W^+$ production (right) at RHIC.} \label{figxeta}  }
\end{figure}

We next investigate the kinematics of $W$ production at RHIC. 
In our view, it is preferable to consider distributions in 
lepton {\it rapidity}, rather than transverse momentum, since
there is a particularly strong and direct correlation between lepton 
rapidity and the partonic momentum fractions. This correlation was 
already evident in the LO asymmetry as a function of the $W$'s rapidity 
discussed in Eq.~(\ref{loW}), for which we had $x_{1,2}=\frac{M_W}
{\sqrt{S}} e^{\pm y_W}$. One can expect that, at least to some extent, 
this relation between momentum fractions and rapidity at the gauge 
boson level will be inherited by the lepton. Figure~\ref{figxeta} shows
the correlation between the averages of the momentum fractions, $\langle x_{1,2} \rangle$, and the rapidity of the charged lepton computed at NLO accuracy 
for $W^-$ (left-hand side) and $W^+$ production (right-hand side) in 
spin-averaged collisions\footnote{The correlation remains the same 
when the $Z/\gamma$ contribution is included.}.
A remarkably strong correlation is found between $\langle x_{1,2}\rangle$ 
and $\eta_l$ in both cases. Large {\it negative} lepton rapidity corresponds to {\it small} ({\it large}) momentum fractions $x_1$ ($x_2$). The opposite occurs for large positive rapidities. Actually, as a rough approximation one can parameterize these 
correlations by the simple "empirical" formulas
\begin{equation}
\langle x_{1,2}\rangle \sim \frac{M_W}{\sqrt{S}} e^{\pm \eta_l/2}\;.
 \end{equation}
Considering that RHIC experiments will allow one to reach rapidities of the order of 
$|\eta_l|\sim 2$, one can expect sensitivity to the polarized quark and anti-quark distributions in the region $0.05 \lesssim x \lesssim 0.4$.
We note that similar results as in Fig.~\ref{figxeta} were also found in
Ref.~\cite{NadYuan2}.

%%%%%%%%%%%%%%%%%%%%%%%%%%%%%%%%%%%%%%%%%%%%%%%%%%%%%%%%%%%%%%%%%%%%%%%%%%
%%%%%%%%%%%%%%%%%%%%%%%%%%%%%%%%%%%%%%%%%%%%%%%%%%%%%%%%%%%%%%%%%%%%%%%%%%%%%%%%

Because of the correlation shown in Fig.~\ref{figxeta}, the combinations 
of parton distributions predominantly probed will vary with $\eta_l$.
However, here also the underlying structure of the weak interactions enters. 
For $W^-$ production, neglecting all partonic processes but the 
dominant $\bar{u}d \to W^-\to e^-\bar{\nu}_e$ 
one, the asymmetry is found to be 
given by 
\begin{equation}
A_L^{e^-}\approx\frac{\int_{\otimes (x_1,x_2)} \left[
\Delta \bar{u}(x_1)d(x_2)(1-\cos\theta)^2-
\Delta d(x_1) \bar{u}(x_2)(1+\cos\theta)^2\right]}
{\int_{\otimes (x_1,x_2)} \left[\bar{u}(x_1)d(x_2)(1-\cos\theta)^2+
d(x_1) \bar{u}(x_2)(1+\cos\theta)^2\right]}\, ,
\label{eq:w-lo}
\end{equation} 
where $\int_{\otimes (x_1,x_2)}$ denotes an appropriate 
convolution over momentum fractions, and 
where $\theta$ is the polar angle of the electron in the partonic
c.m.s., with $\theta>0$ in the forward direction of
the polarized parton. Note that $\theta$ itself depends 
on the momentum fractions and on the lepton's rapidity. 
At large negative $\eta_l$, 
one has $x_2\gg x_1$ and $\theta\sim\pi$. In this case, the 
first terms in the numerator and denominator of 
Eq.~(\ref{eq:w-lo}) strongly dominate, since 
the combination of parton distributions, $\Delta\bar{u}(x_1)
d(x_2)$, and the angular factor, $(1-\cos\theta)^2$, each dominate 
over their counterpart in the second term. Therefore, the asymmetry 
provides a clean probe of $\Delta\bar{u}(x_1)/\bar{u}(x_1)$ at medium
values of $x_1$. By similar reasoning, at forward rapidity $\eta_l\gg 0$ the 
second terms in the numerator and denominator of 
Eq.~(\ref{eq:w-lo}) dominate, giving access
to $-\Delta d(x_1)/d(x_1)$ at relatively high $x_1$.
For the $W^+$ production channel one has instead of~(\ref{eq:w-lo})
\begin{equation}
A_L^{e^+}\approx\frac{\int_{\otimes (x_1,x_2)} \left[
\Delta \bar{d}(x_1)u(x_2)(1+\cos\theta)^2-
\Delta u(x_1) \bar{d}(x_2)(1-\cos\theta)^2\right]}
{\int_{\otimes (x_1,x_2)} \left[\bar{d}(x_1)u(x_2)(1+\cos\theta)^2+
u(x_1) \bar{d}(x_2)(1-\cos\theta)^2\right]}\,  .
\label{eq:w+lo}
\end{equation} 
Here the distinction of the two contributions by considering 
large negative or positive lepton rapidities is less clear-cut than
in the case of $W^-$. For example, at negative $\eta_l$
the partonic combination $\bar{d}(x_1)u(x_2)$ will dominate, but
at the same time $\theta\sim\pi$ so that the angular 
factor $(1+\cos\theta)^2$ given by the basic electroweak interaction
is small. Likewise, at positive 
$\eta_l$ the dominant partonic combination 
$\Delta u(x_1) \bar{d}(x_2)$ is suppressed by the angular factor
because $\theta\sim 0$.
So both terms in Eq.~(\ref{eq:w+lo}) will compete essentially 
for all $\eta_l$ of interest. This is also the reason why the $W^-$
cross section can become larger than the $W^+$ one at high
rapidities (see Fig.~\ref{figWZunpol}). As was discussed
in Refs.~\cite{Whad1,Whad2}, the study of hadronic $W$ decays, 
in particular of charmed final states~\cite{Whad1}, might help 
in improving this situation.

\begin {figure}[!ht]
\begin{center}
\includegraphics[width = 3.2in]{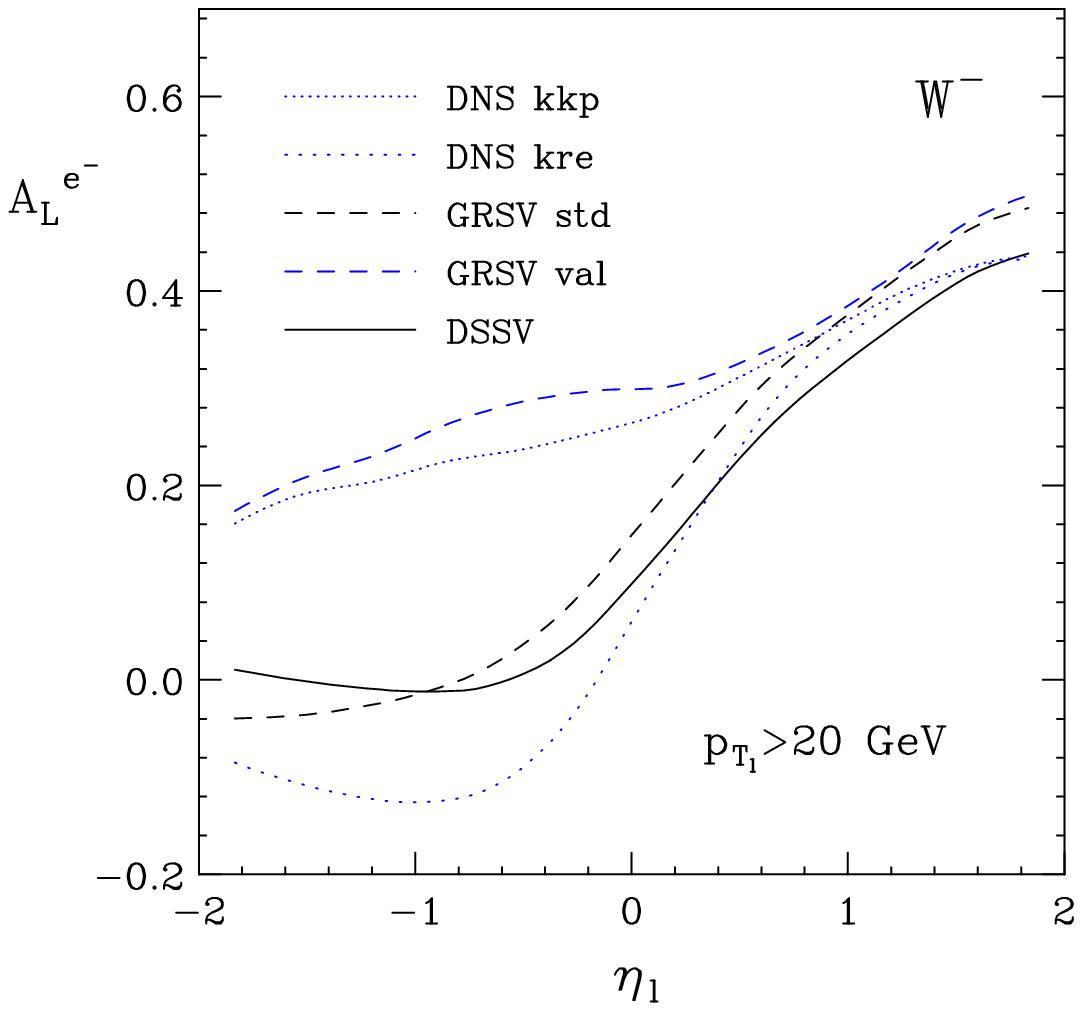} 
\hspace*{-0.cm}
\includegraphics[width = 3.2in]{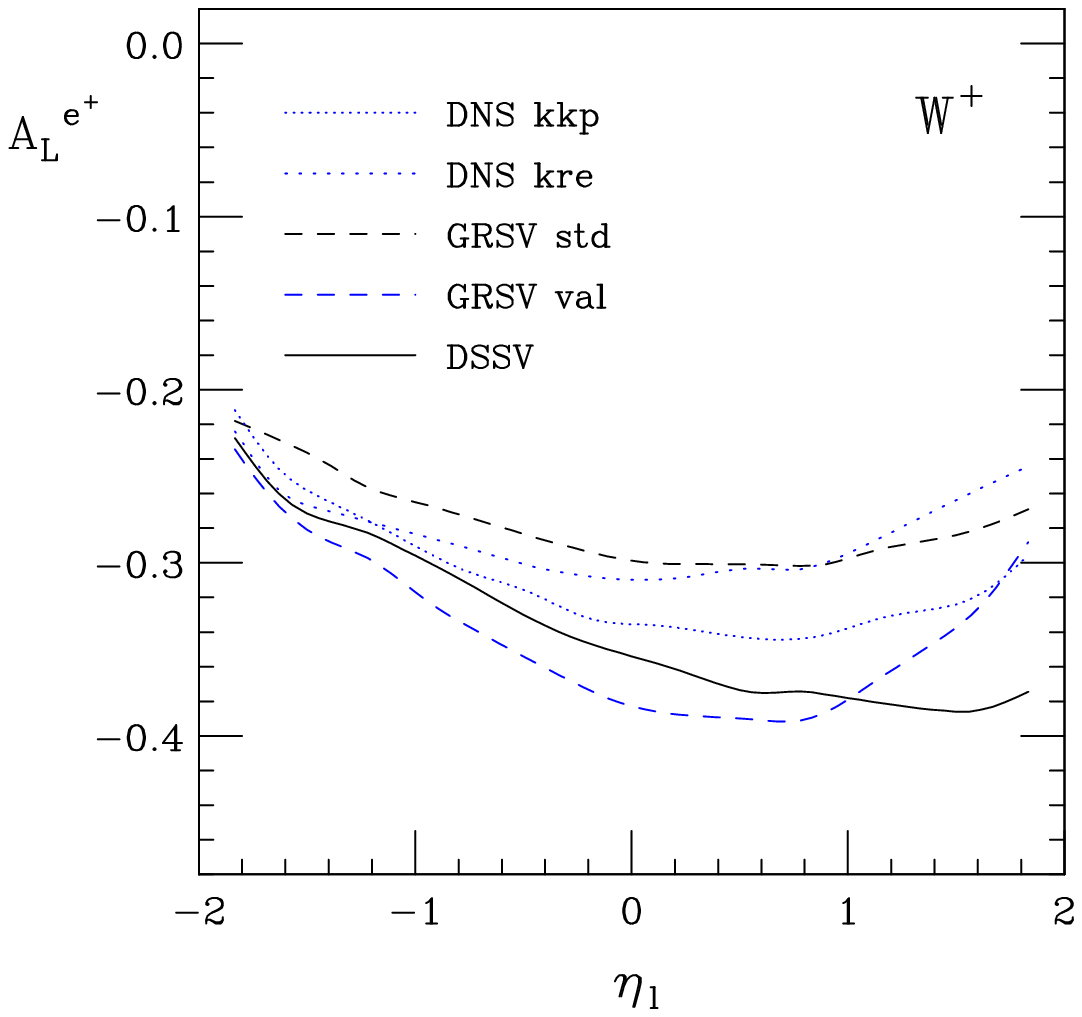} 
\end{center}
\caption{{\it Rapidity dependence of the NLO single-spin asymmetries 
$A_L^{e^-}$ for electrons and $A_L^{e^+}$ for positrons at RHIC, for
the various sets of polarized parton distribution functions shown 
in Fig.~\ref{figpdf}. Only leptons produced by $W^\pm$ boson exchange are considered here.} \label{figw-}  }
\end{figure}

The features displayed by Eqs.~(\ref{eq:w-lo}),(\ref{eq:w+lo}) are 
fully reflected in the behavior of the calculated NLO spin asymmetries.
Figure~\ref{figw-} shows $A_L^{e^-}$ for electrons and 
$A_L^{e^+}$ for positrons at RHIC, as functions of the charged lepton's rapidity, considering only leptons arising from $W^\pm$ boson exchange.
We are now using all of the sets of polarized parton distributions that we 
introduced earlier. The spread in the predictions for the asymmetry $A_L^{e^-}$
at $\eta_l\lesssim 0$ directly  reflects the dispersion in both the absolute magnitude and sign of the different $\Delta\bar{u}(x)$ distributions shown in Fig.~\ref{figpdf} in the range $ 0.05 \lesssim x\lesssim 0.2$. On the other hand, the asymmetry becomes large and positive at high $\eta_l$, which reflects the fact that 
$\Delta d(x)$ remains negative at high $x$ for all sets of polarized parton 
distributions considered here. $A_L^{e^+}$ does not show as clear features,
for the reasons we just discussed. Nevertheless, at $\eta_l\gtrsim 0$
one can observe again that the spread of the predictions for the asymmetry is 
quite strongly correlated to the one found for the $\Delta\bar{d}$ distributions 
at $0.15 \lesssim x\lesssim 0.3$ in Fig.~\ref{figpdf}. Overall, the 
asymmetry is negative because of the contribution from $\Delta u$ in
Eq.~(\ref{eq:w+lo}), which is known to be positive from lepton scattering.

%%%%%%%%%%%%%%%%%%%%%%%%%%%%%%%%%%%%%%%%%%%%%%%%%%%%%%%%%%%%%%%%%%%%%%%%%%
%%%%%%%%%%%%%%%%%%%%%%%%%%%%%%%%%%%%%%%%%%%%%%%%%%%%%%%%%%%%%%%%%%%%%%%%%%%%%%%%

It is worth pointing out that in Ref.~\cite{dssvprd} spin asymmetries for 
the same sets of parton distributions as in Fig.~\ref{figw-} were shown, 
but at LO. Our NLO results turn out to be 
very close to the LO ones of~\cite{dssvprd}. This is easily understood
because the bulk of the NLO corrections in the $q\bar{q}'$ channel
is the same in the spin-averaged and the polarized cases, so that the
corrections cancel to a high degree. This result was also observed in
the study~\cite{asmita} of the $W$ cross section without the leptonic
decay. We do stress, however, that the individual cross sections in the
numerator and the denominator of $A_L$ receive significant NLO corrections of
${\cal O}(30\%)$.

\begin {figure}[!ht]
\begin{center}
\includegraphics[width = 3.2in]{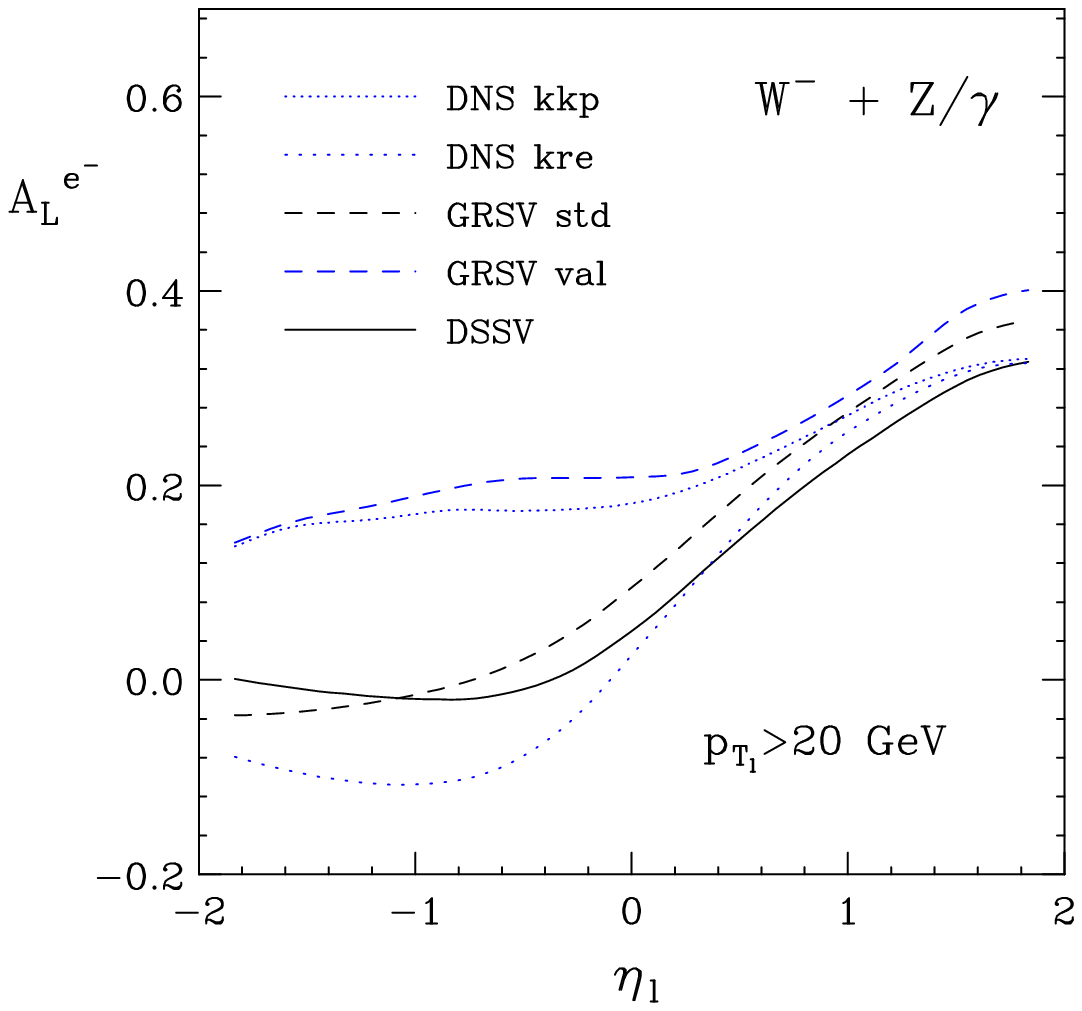} 
\hspace*{-0.cm}
\includegraphics[width = 3.2in]{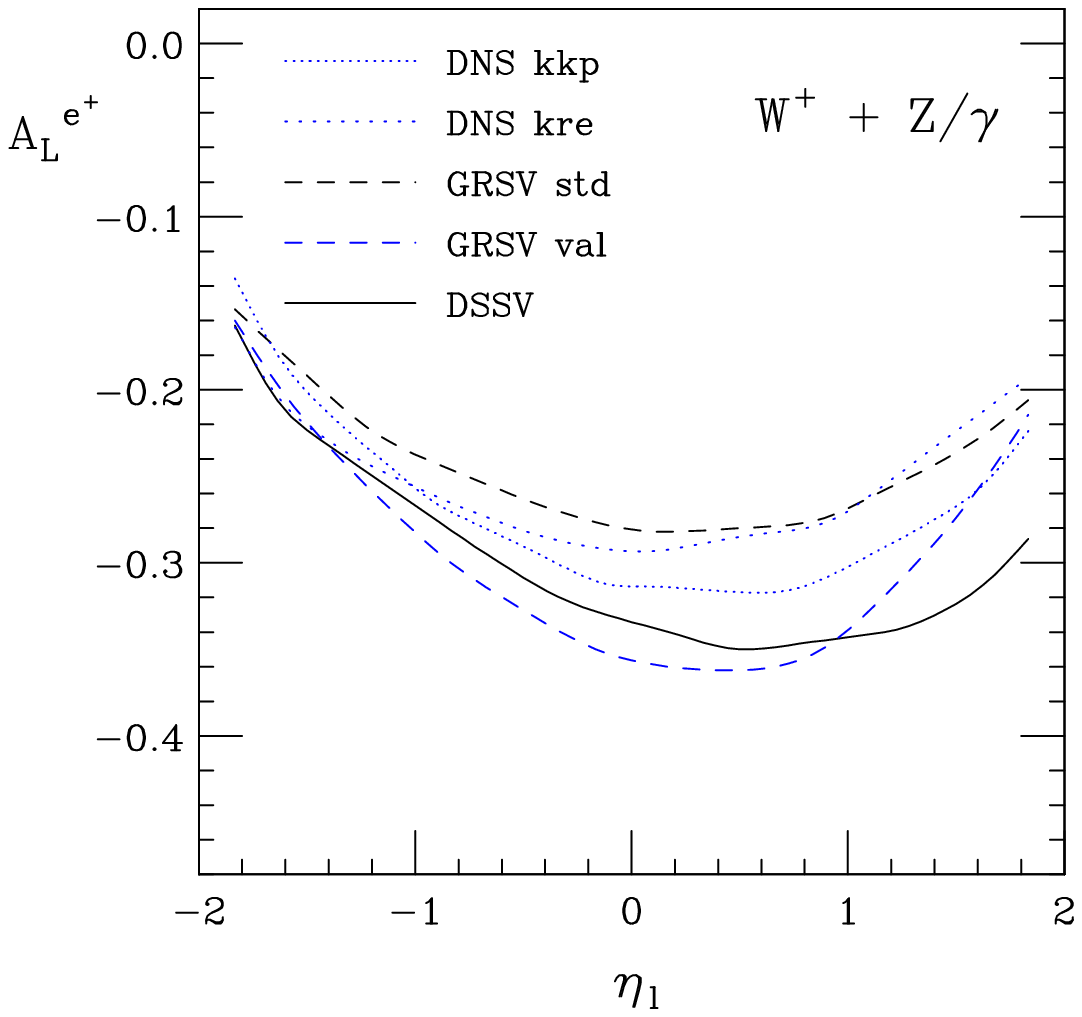} 
\end{center}
\caption{{\it Same as in Fig.~\ref{figw-} but including also the contribution 
from $Z/\gamma$ exchange.} \label{figw}  }
\end{figure}

For completeness, we show in Fig.~\ref{figw} the asymmetries
computed by including also leptons produced by $Z/\gamma$ boson exchange. As expected, the inclusion of `background' leptons results in a reduction of the asymmetry due to the increase in the unpolarized cross section. Consistently 
with the results presented in Fig.~\ref{figWZunpol}, the effect is more noticeable at larger rapidities.

%%%%%%%%%%%%%%%%%%%%%%%%%%%%%%%%%%%%%%%%%%%%%%%%%%%%%%%%%%%%%%%%%%%%%%%%%%
%%%%%%%%%%%%%%%%%%%%%%%%%%%%%%%%%%%%%%%%%%%%%%%%%%%%%%%%%%%%%%%%%%%%%%%%%%%%%%%%

\begin {figure}[!ht]
\begin{center}
\includegraphics[width = 5in]{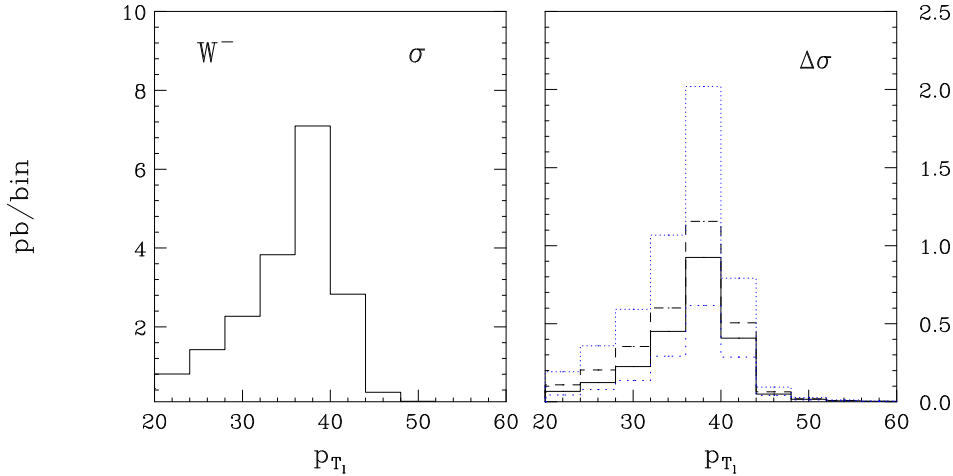} 
\end{center}%}
\caption{{\it Left: transverse momentum dependence of the unpolarized
cross section for electron production at RHIC. Right: same for the 
single-spin cross section. Here the lines follow the same pattern as 
in Fig.~\ref{figw-}. } \label{figw-pt}  }
\end{figure}

Figure~\ref{figw-pt} shows the 
spin-averaged and single-spin cross sections for electron production,
as functions of $p_{T_l}$, integrated over $|\eta_l| \leq 1$. 
As expected, a peak is found around $p_{T_l}\sim M_W/2$ in both cases. 
For the single-polarized cross section, we are again using various sets
of polarized parton distributions. One can see that the dependence on the 
polarized distribution functions is apparent in the magnitude of the 
cross section, but hardly in the shape of the transverse momentum distribution.
The latter is mainly determined by the properties of the gauge boson (like its mass and width) and general features of QCD radiation and, therefore, is very similar 
for both unpolarized and polarized cross-sections. 
In other words, integration over a significant region of rapidity
washes out many of the features that we found in Fig.~\ref{figxeta}
for the polarized parton distributions. While this situation may possibly be
improved by integrating over non-central regions of lepton rapidity,
our studies overall confirm our expectation that in order to improve our 
knowledge of the polarized anti-quark distributions it is preferable 
to study the lepton rapidity dependence of the asymmetry instead of its
transverse momentum one.

%%%%%%%%%%%%%%%%%%%%%%%%%%%%%%%%%%%%%%%%%%%%%%%%%%%%%%%%%%%%%%%%%%%%%%%%%%
%%%%%%%%%%%%%%%%%%%%%%%%%%%%%%%%%%%%%%%%%%%%%%%%%%%%%%%%%%%%%%%%%%%%%%%%%%%%%%%%

Apart from providing a good correlation between momentum fractions and
lepton rapidity, another advantage of rapidity distributions 
integrated over a wide region of transverse momentum is that
these are insensitive to soft-gluon and resummation effects. 
As discussed in the Introduction,
the distribution near $p_{T_l}=M_W/2$ is sensitive to soft-gluon 
effects. Proper inclusion of these effects in the theoretical description
plays a role, for example, in determinations of the $W$ mass
from the lepton $p_{T_l}$ distributions at the Tevatron~\cite{tev},
or in transverse-spin studies in $W$ boson production
at RHIC~\cite{Kang:2009bp}.
The RHICBOS code~\cite{NadYuan1,NadYuan2} includes an all-order
resummation of the next-to-leading logarithms in the $W$ 
transverse momentum $q_T$, which are relevant near $p_{T_l}=M_W/2$. 
However, once
one integrates over a sufficiently large region of lepton transverse momentum,
these logarithms turn into finite higher-order (beyond NLO) corrections to the 
transverse-momentum distribution, and neither is their resummation
necessary nor is it guaranteed to provide an improved theoretical 
description, as there can and will be many other corrections
of similar size that are not taken into account. From the point of view
of extracting polarized parton distribution functions, it therefore 
seems advisable to us
to focus on observables integrated over the lepton's transverse momentum,
because these are insensitive to soft-gluon effects, and to use a plain NLO 
calculation. Thanks to the fact that the $W$ mass sets a very large
scale so that the strong coupling is small, and because quark anti-quark
annihilation is the dominant partonic channel, any "partial" beyond-NLO and
non-perturbative effects remaining from $q_T$-resummation are
expected to be relatively small for such observables. Comparing
the total $W^+$($W^-$) spin-averaged cross sections for $p_{T,l}>25$~GeV,
$|\eta_l|<1$ and setting the scales to $\mu_F=\mu_R=M_W$, we find $75.4$~pb  ($17.7$~pb) with our NLO code, while RHICBOS
gives\footnote{We thank B. Surrow for providing these 
numbers.} $80.5$~pb ($18.8$~pb). The difference will be in part due to different
choices for  electroweak parameters and parton distributions,
but also due to the additional effects  in RHICBOS just described.
Somewhat larger differences between the codes occur if one
considers the differential cross section at high lepton rapidity
and, of course, as a function of transverse momentum at
$p_{T,l}\sim M_W/2$. We have checked that these differences do not, however, 
significantly affect any of the previously performed sensitivity
studies by the RHIC experiments.

%%%%%%%%%%%%%%%%%%%%%%%%%%%%%%%%%%%%%%%%%%%%%%%%%%%%%%%%%%%%%%%%%%%%%%%%%
%%%%%%%%%%%%%%%%%%%%%%%%%%%  Section Expectations  %%%%%%%%%%%%%%%%%%%%%
%%%%%%%%%%%%%%%%%%%%%%%%%%%%%%%%%%%%%%%%%%%%%%%%%%%%%%%%%%%%%%%%%%%%%%%%%
\section{Toy Analysis of $W$ Spin Asymmetries in Terms of Polarized Parton
Distributions \label{secIV}}

\begin {figure}[!ht]
\begin{center}
\includegraphics[width = 5.0in]{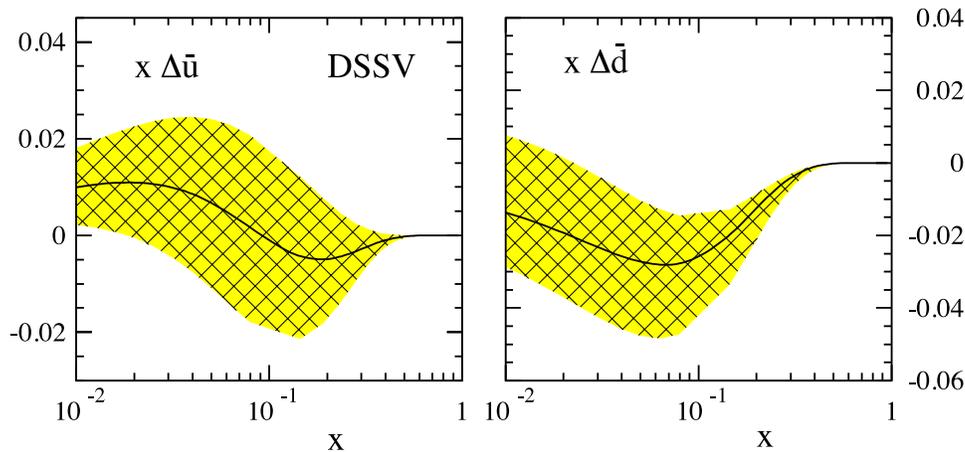} 
\end{center}%}
\caption{{\it $\Delta \bar{u}(x,Q^2=10\, {\rm GeV}^2)$ and $\Delta \bar{d}(x,Q^2=10\, {\rm GeV}^2)$ as obtained in the DSSV 
analysis~\cite{dssvprl,dssvprd}. The bands correspond to changes of $2\%$  
in the total $\chi^2$ of the fit, as discussed in the DSSV paper.} \label{dssv}  }
\end{figure}

While the results presented in the previous section indicate a strong sensitivity of 
the single-spin asymmetries to the polarized light quark and anti-quark distributions, 
it is quite difficult to quantify from them the impact future RHIC measurements will 
have. In order to investigate this, we perform a more detailed 
analysis. Our strategy is to "simulate" a set of RHIC data under hopefully 
realistic conditions, and to add this set to the data sets included in the 
published DSSV~\cite{dssvprd,dssvprl} global analysis. A new fit is then 
performed, for which the simulated data set is included, and the impact
of future $W$ data from RHIC is gauged from the improvement found
in this fit for the extracted polarized distributions. As discussed
in the Introduction, knowledge about the sea quark and anti-quark
polarizations in the nucleon so far comes entirely from the SIDIS 
spin asymmetries. In Fig.~\ref{dssv} we recall the DSSV results for
the polarized anti-quark distributions, including their respective uncertainty 
bands, which were obtained in DSSV by performing a Lagrangian multiplier
analysis of the truncated moments $\int_{0.001}^1 dx\, \Delta \bar{u}(x,Q^2=10\, {\rm GeV}^2)$ and $\int_{0.001}^1 dx\, \Delta \bar{d}(x,Q^2=10\, {\rm GeV}^2)$, and allowing modifications of $2\%$ in the total $\chi^2$ of the fit. 

%%%%%%%%%%%%%%%%%%%%%%%%%%%%%%%%%%%%%%%%%%%%%%%%%%%%%%%%%%%%%%%%%%%%%%%%%%
%%%%%%%%%%%%%%%%%%%%%%%%%%%%%%%%%%%%%%%%%%%%%%%%%%%%%%%%%%%%%%%%%%%%%%%%%%%%%%%%

\begin {figure}[!ht]
\begin{center}
\includegraphics[width = 4.50in]{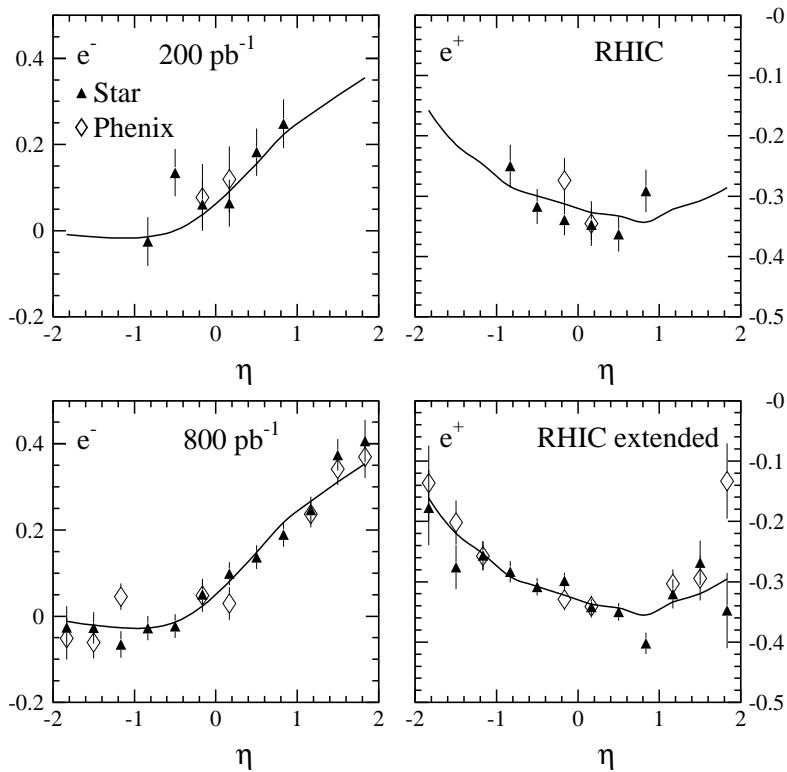} 
\end{center}%}
\caption{{\it   Simulated data generated for $e^-$ (left) and  $e^+$ (right) 
single-spin asymmetries at STAR and Phenix for two possible extreme scenarios: 
integrated luminosity ${\cal L}= 200\, {\rm pb}^{-1}$ and present RHIC
rapidity coverage (upper row), and integrated luminosity ${\cal L}= 800\, 
{\rm pb}^{-1}$ and upgraded rapidity coverage as described in the text
(lower row). The "simulated data points" have been estimated by performing
a NLO calculation with the DSSV set of polarized parton distributions,
followed by a Gaussian shift of the points. The solid lines represent the 
expectation from the DSSV set.} \label{wdata}}
\end{figure}

In order to generate a simulated "RHIC data set" we proceed as follows: we
compute the NLO single-longitudinal spin asymmetries $A_L^{e^+}$ and 
$A_L^{e^-}$ using the central DSSV set of polarized parton distributions \footnote{Asymmetries are evaluated including both $W$ and $Z/\gamma$ boson exchange contributions}.
We then randomly shift the calculated asymmetries, assuming a Gaussian
distribution of their statistical uncertainties. The latter are estimated using the
usual formula $\delta A_L=1/(P\sqrt{{\cal L}\sigma})$. We
assume a polarization of $P=60\%$. For the integrated luminosity 
we consider two values: 
${\cal L}= 200\, {\rm pb}^{-1}$ and  ${\cal L}= 800\, {\rm pb}^{-1}$.
Concerning rapidity coverage, we focus first on the present coverage for
the Phenix ($|\eta_l|<0.35$) and STAR ($|\eta_l|<1$) experiments. Later, we 
also investigate the impact of extended rapidity coverage as given by 
$|\eta_l|<0.35$ and $1<|\eta_l|<2$ for Phenix\footnote{The upgrade planned for Phenix will actually extend the rapidity coverage to 
$1.2 \lesssim |\eta_l|\lesssim 2.4$} and  $|\eta_l|<2$ for STAR . 
The generated pseudo-data points are shown in Fig.~\ref{wdata}
for the two extreme scenarios (smaller luminosity and rapidity coverage 
vs large luminosity and rapidity coverage), along with the predictions 
according to the DSSV set of polarized parton distributions. We have
assumed bin sizes in rapidity of $\Delta \eta_l=0.33$, so that there
are six data points in $|\eta_l|\leq 1$. 

%%%%%%%%%%%%%%%%%%%%%%%%%%%%%%%%%%%%%%%%%%%%%%%%%%%%%%%%%%%%%%%%%%%%%%%%%%
%%%%%%%%%%%%%%%%%%%%%%%%%%%%%%%%%%%%%%%%%%%%%%%%%%%%%%%%%%%%%%%%%%%%%%%%%%%%%%%%

\begin {figure}[!ht]
\begin{center}
\includegraphics[width = 5.0in]{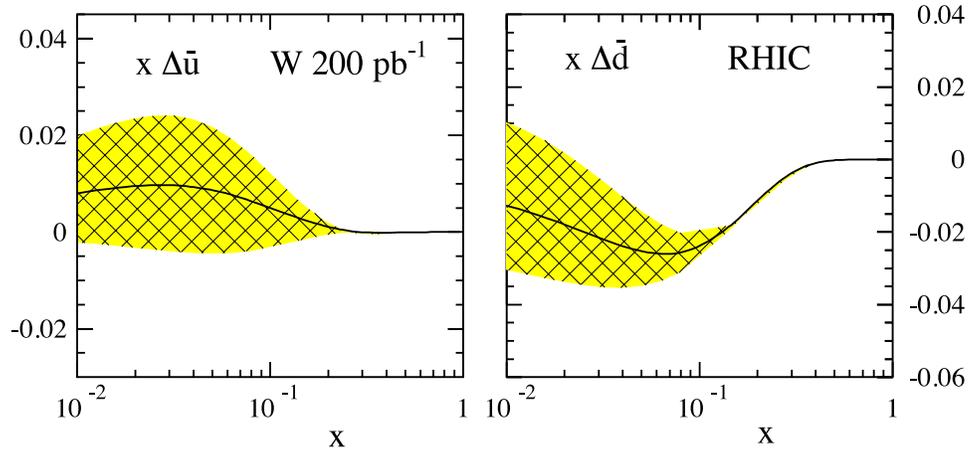} 
\end{center}%}
\caption{{\it Result of a global fit performed by including the simulated data 
generated with the lower luminosity and smaller rapidity coverage scenario 
(upper row in Fig.~\ref{wdata}).} \label{w1}  }
\end{figure}

\begin {figure}[!ht]
\begin{center}
\includegraphics[width = 5.0in]{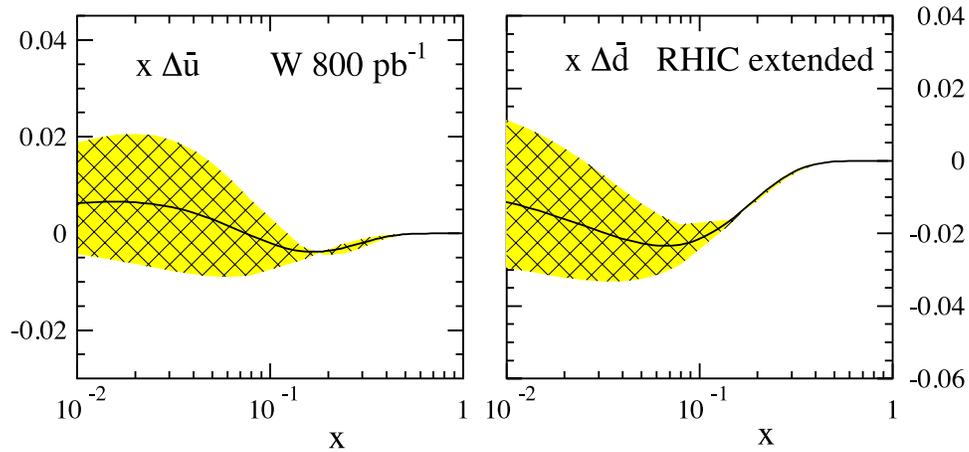} 
\end{center}%}
\caption{{\it Result of a global fit performed by including the simulated data 
generated with the larger luminosity and rapidity coverage scenario (lower row in Fig.~\ref{wdata}).} \label{w2}  }
\end{figure}

In order to perform the actual fit, we produce the pre-calculated grids
as described in Sec.~\ref{secII}. The choice of bins $\Delta \eta_l=0.33$
means that we can use the same grids for both RHIC experiments,
because the rapidity range covered for Phenix corresponds in good 
approximation to two bins with $\Delta \eta_l=0.33$.
In our first fit, we include in the DSSV analysis the pseudo-data generated with 
the lower luminosity and the rapidity coverage presently available at
RHIC. The outcome of this global fit is shown for the polarized
anti-quark distributions in Fig.~\ref{w1}, including their resulting
$\Delta \chi^2/\chi^2=2\%$ 
uncertainties determined in the same way as in the published DSSV 
analysis. By comparing to Fig.~\ref{dssv}, one observes that 
there is little modification of the actual distributions, as compared to the 
original DSSV ones, but a clear reduction in their uncertainty bands. 
This effect turns out to be very noticeable for $x\gtrsim 0.1$, as expected 
considering the rapidity coverage of the pseudo-data added to the global fit. 
The decrease in the uncertainty band is also more noticeable in case of 
$\Delta \bar{u}$, confirming the larger sensitivity of $e^-$ asymmetries 
to this distribution. At values of $x\sim 0.01$ there is almost no change in 
the distributions and their uncertainties, since the single-spin asymmetries at 
RHIC are rather insensitive to such values of $x$. The lower row of 
Fig.~\ref{w2} shows the result of the corresponding fit for the larger
luminosity/rapidity coverage scenario. Here the impact of RHIC data can be observed 
down to values of $x\sim 0.05$, thanks to the extended coverage in 
$\eta_l$. 

%%%%%%%%%%%%%%%%%%%%%%%%%%%%%%%%%%%%%%%%%%%%%%%%%%%%%%%%%%%%%%%%%%%%%%%%%%
%%%%%%%%%%%%%%%%%%%%%%%%%%%%%%%%%%%%%%%%%%%%%%%%%%%%%%%%%%%%%%%%%%%%%%%%%%%%%%%%

%%%%%%%%%%%%%%%%%%%%%%%%%%%%%%%%%%%%%%%%%%%%%%%%%%%%%%%%%%%%%%%%%%%%%%%%%%
%%%%%%%%%%%%%%%%%%%%%%%%%%%%%%%%%%%%%%%%%%%%%%%%%%%%%%%%%%%%%%%%%%%%%%%%%%%%%%%%

\begin {figure}[!ht]
\begin{center}
\includegraphics[width = 5.0in]{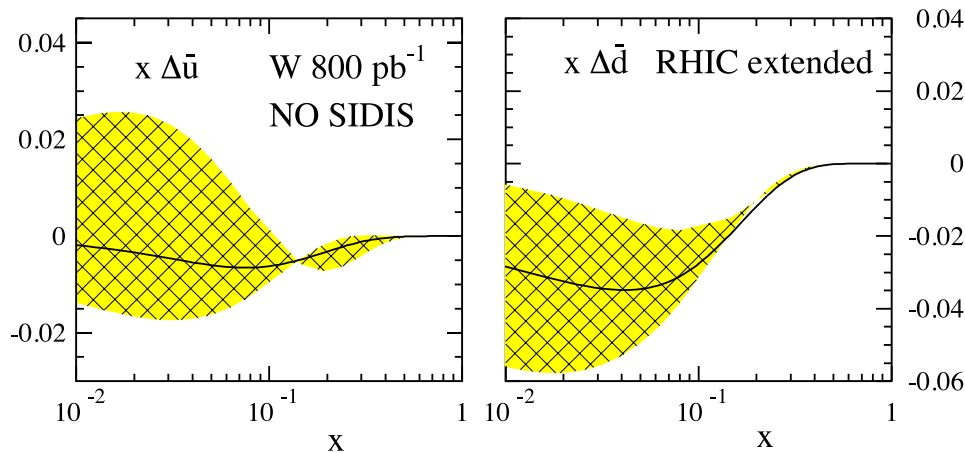} 
\end{center}%}
\caption{{\it Same as Fig.~\ref{w2}, but excluding all SIDIS data from the fit.} 
\label{w3}  }
\end{figure}

The pseudo-data used in this analysis were generated to be in full agreement, 
within statistical errors, with the expectation from the DSSV set. Since the 
latter provides an excellent description of the available SIDIS data, we have
effectively assumed that constraints on the polarized parton distributions 
emerging from SIDIS and from $W$ production at RHIC are in agreement.
This should be the case, of course, if both are described by factorized
perturbative QCD at leading-power. From a theoretical point of view,
$W$ production provides the more reliable source of information, so 
if any discrepancy between SIDIS and $W$ production were found,
it would likely point to higher-twist contributions in SIDIS, or ill-understood 
issues related to fragmentation. In this context, it is interesting to ask 
what the impact of future RHIC data would be if all SIDIS data were
removed from the global fit. We have performed such an analysis
for the scenario with larger luminosity and rapidity coverage. 
The result is shown in Fig.~\ref{w3}. We find that for $x\gtrsim 0.07$ 
the simulated $W$ data put a somewhat better constraint on 
$\Delta \bar{u}$ and $\Delta \bar{d}$ than the SIDIS data presently 
do~\footnote{We remind the reader that the new preliminary COMPASS
SIDIS data~\cite{ref:compass-sidisnew} were not yet included in the 
DSSV analysis~\cite{dssvprl,dssvprd} 
and are hence not included in the present study.}.
Toward smaller $x$, the distributions are of course only very 
loosely determined because the $W$ spin asymmetries are not 
sensitive to this region. All in all, there are very good prospects 
for a much better determination of the polarized anti-quark distributions
from RHIC and SIDIS measurements.

%%%%%%%%%%%%%%%%%%%%%%%%%%%%%%%%%%%%%%%%%%%%%%%%%%%%%%%%%%%%%%%%%%%%%%%%%%
%%%%%%%%%%%%%%%%%%%%%%%%%%%%%%%%%%%%%%%%%%%%%%%%%%%%%%%%%%%%%%%%%%%%%%%%%%%%%%%%

We end by stressing that there are numerous experimental issues 
(like efficiencies for lepton detection, correct background subtraction, 
other systematic uncertainties, etc.) that have not been included in 
this simple analysis and that would tend to decrease the impact of the 
real data in the global fit. We regard this study as a "proof-of-principle"
that shows that RHIC $W$ asymmetry data can be straightforwardly
included in a global analysis of polarized parton distributions.
Future, more detailed, studies will need to be carried out. 

%%%%%%%%%%%%%%%%%%%%%%%%%%%%%%%%%%%%%%%%%%%%%%%%%%%%%%%%%%%%%%%%%%%%%%%%%
%%%%%%%%%%%%%%%%%%%%%%%%%%%  Section 5 Conclusions  %%%%%%%%%%%%%%%%%%%%%
%%%%%%%%%%%%%%%%%%%%%%%%%%%%%%%%%%%%%%%%%%%%%%%%%%%%%%%%%%%%%%%%%%%%%%%%%
\section{Conclusions  \label{secV}}

We have presented a new next-to-leading order calculation 
of the cross section and longitudinal spin asymmetry 
for the process $pp\to \ell^{\pm} X$ at RHIC, through an 
intermediate electroweak gauge boson. The spin-asymmetry 
is the main probe of the light quark and anti-quark helicity 
distributions at RHIC. We have developed a multi-purpose 
Monte-Carlo integration program. Our code has the advantage
that it allows one to directly include forthcoming RHIC data into 
a global analysis of spin-dependent parton densities, using 
the Mellin technique of~\cite{sv,dssvprl,dssvprd}. Compared to 
the RHICBOS code~\cite{NadYuan1,NadYuan2}, our program 
does not include any soft-gluon $q_T$-resummation effects,
as we advocate the use of observables at RHIC that are insensitive
to such effects. In particular, we have emphasized the advantage
of the lepton rapidity distribution over the transverse-momentum
one.

Our phenomenological results indicate a good sensitivity of
the single-longitudinal spin asymmetries at RHIC to the 
light quark and anti-quark helicity distributions. This
finding is in line with those of previous 
studies~\cite{spinplan,NadYuan1,NadYuan2,dssvprd}. 
Contributions from $Z$ exchange are found to be generally
non-negligible. As a benchmark application of our program,
we have performed a toy global analysis of "simulated" RHIC
spin asymmetry data along with the present lepton scattering
and RHIC high-$p_T$ jet and hadron data. We find that
RHIC has a great potential for providing better constraints
on the light quark and anti-quark helicity distributions, 
in particular at moderately large momentum fractions.
Once precise data becomes available from RHIC, the 
consistency and interplay with constraints from SIDIS will 
be particularly interesting to investigate. While more refined 
sensitivity studies will be needed, we regard our findings as
a very encouraging signal that precise information on the
nucleon's polarized light quark and anti-quark distributions
will become available before too long. Such information will
likely offer important insights into the inner structure of the nucleon
and the dynamics of QCD.

%%%%%%%%%%%%%%%%%%%%%%%%%%%%%%%%%%%%%%%%%%%%%%%%%%%%%%%%%%%%%%%%%%%%%%%%%
%%%%%%%%%%%%%%%%%%%%%%%%%%%%%%%%%   Acknowledgments   %%%%%%%%%%%%%%%%%%%
%%%%%%%%%%%%%%%%%%%%%%%%%%%%%%%%%%%%%%%%%%%%%%%%%%%%%%%%%%%%%%%%%%%%%%%%%
\section*{Acknowledgments}
We are grateful to J.\ Haggerty, J.W.\ Qiu, R.\ Seidl and B.\ Surrow for useful discussions.
The work of D.deF. has been partially supported by Conicet,
 UBACyT, ANPCyT and the Guggenheim Foundation.
W.V.'s work has been supported by the U.S.\ Department of Energy
(contract number DE-AC02-98CH10886) and by LDRD 
project 08-004 of Brookhaven National Laboratory.

\end{document}